\documentclass[a4paper,12pt]{article}
\usepackage{amsmath,amsfonts,amssymb,cite}
\usepackage{graphicx
}

\begin{document}

\begin{center}
{\LARGE\bf Note on universal description of heavy and light
mesons}
\end{center}

\begin{center}
{\large S. S. Afonin and I. V. Pusenkov}
\end{center}

\begin{center}
{\small V. A. Fock Department of Theoretical Physics,
Saint-Petersburg
State University, 1 ul. Ulyanovskaya, St. Petersburg, 198504, Russia\\
}
\end{center}

\begin{abstract}
The experimental spectrum of excited $S$-wave vector mesons with
hidden quark flavor reveals a remarkable property: For all
flavors, it is approximately linear in mass squared, $m_n^2\approx
a(n+b)$, $n$ is the radial quantum number. We draw attention to
the fact that such a universal behavior for any quark mass cannot
be obtained in a natural way within the usual semirelativistic
potential and string-like models --- if the Regge-like behavior is
reproduced for the mesons composed of the light quarks, the
trajectories become essentially nonlinear for the heavy-quark
sector. In reality, however, the linearity for the heavy mesons
appears to be even better than for the light ones. In addition,
the slope $a$ is quite different for different quark flavors. This
difference is difficult to understand within the QCD string
approach since the slope measures the interaction strength among
quarks. We propose a simple way for reparametrization of the
vector spectrum in terms of quark masses and universal slope and
intercept. Our model-independent analysis suggests that the quarks
of any mass should be regarded as static sources inside mesons
while the interaction between quarks is substantially
relativistic.
\end{abstract}

\section{Introduction}

The hadron spectroscopy continues to play the major role in the
study of the strong interactions. The main goal is the exhaustive
description of the hadron spectrum in terms of a dimensional
parameter and the quark masses. This task has still not been
solved starting from the QCD Lagrangian. A serious progress was
achieved by the semirelativistic potential
models~\cite{isgur,isgur2,isgur3}). They enjoyed a striking
success in the description of the ground states and related
physics. But their results in the sector of radially excited
hadrons were much more modest.

On the basis of QCD one expects that the formation of resonances
should happen in a universal manner at all available scales. In
the experimental hadron spectroscopy, the most studied sector is
given by the unflavored vector mesons. These resonances have the
quantum numbers of the photon and therefore are intensively
produced in the $e^+e^-$-annihilation which represents a
traditional laboratory for the discovery of new quark flavors and
for precise measurements of the vector spectrum. Since the
creation mechanism for the vector mesons in the
$e^+e^-$-annihilation is identical for any flavor and many
experimental data is available, the case of the unflavored vector
mesons looks the most appropriate for analyzing the manifestations
of universality of the strong interactions in the resonance
formation. These manifestations must be reproduced by any viable
dynamical model.

In the present note, we study the spectroscopic universality among
the vector mesons with hidden flavor --- the $\varphi$, $\psi$,
$\Upsilon$ mesons and their analogues in the sector of $u,d$
quarks, the $\omega$-mesons (the spectrum of $\rho$-mesons is
similar and does not bring anything new in our discussions). Based
on the experimental data, the spectroscopic manifestations of the
universality in question are summarized. We argue that these
manifestations may be explained if the total meson mass is
composed (in the first approximation) of two contributions --- a
non-relativistic one due to the static quark masses and a
relativistic one stemming from the gluon interactions. The mass
spectrum can be described by a simple formula which is universal
for all considered mesons. The found relation is checked and its
parameters are estimated. We indicate then the reason which does
not allow to obtain the proposed relation in the usual potential
and string-like models and speculate on a possible way for
modifying these models.

\begin{figure}[ht]

    \begin{minipage}[ht]{0.46\linewidth}
    \includegraphics[width=1\linewidth]{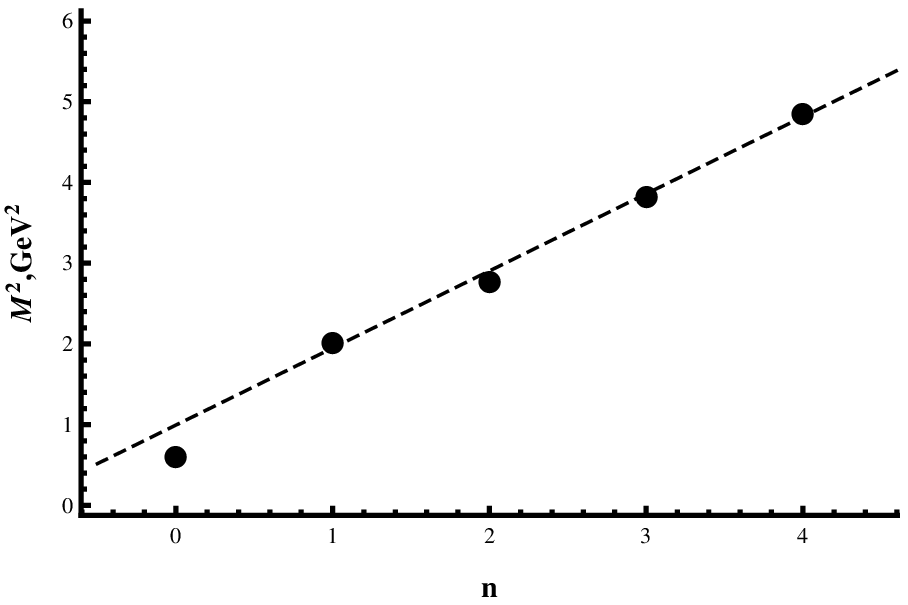} \\
    {\scriptsize Fig. (1a). The spectrum of $\omega$-mesons.
    The experimental points (for this and subsequent figures) are taken from Table 1.}
    \end{minipage}
    \hfill
    \begin{minipage}[ht]{0.46\linewidth}
    \includegraphics[width=1\linewidth]{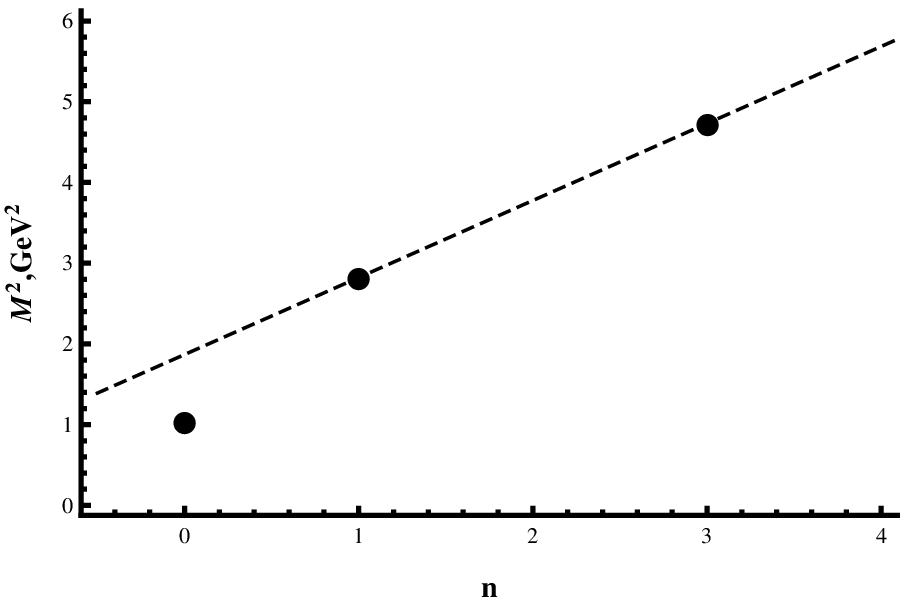} \\
    {\scriptsize Fig. (1b). The spectrum of $\phi$-mesons.\\ \\ \\}
    \end{minipage}

    \vspace{0.7cm}

    \begin{minipage}[ht]{0.46\linewidth}
    \includegraphics[width=1\linewidth]{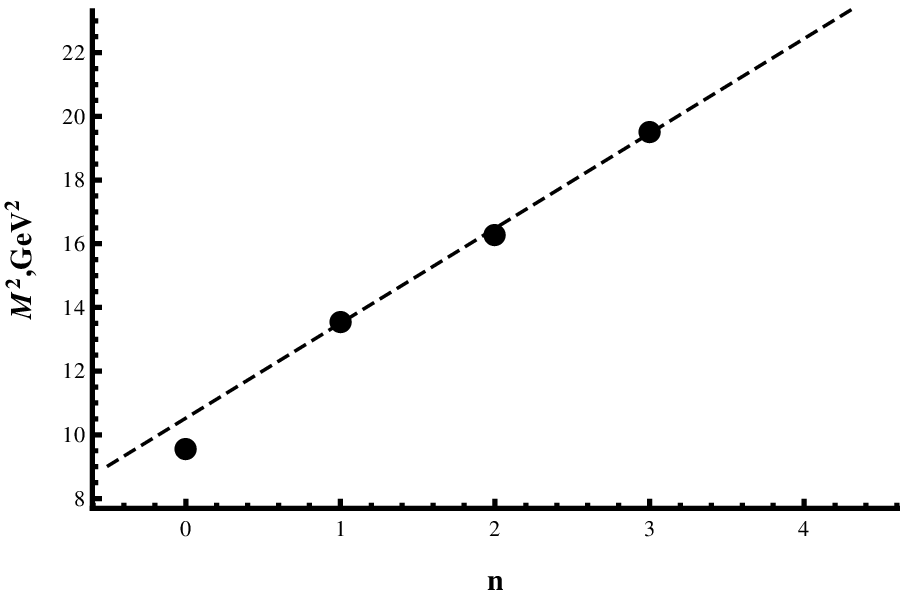} \\
    {\scriptsize Fig. (1c). The spectrum of $\psi$-mesons.}
    \end{minipage}
    \hfill
    \begin{minipage}[ht]{0.46\linewidth}
    \includegraphics[width=1\linewidth]{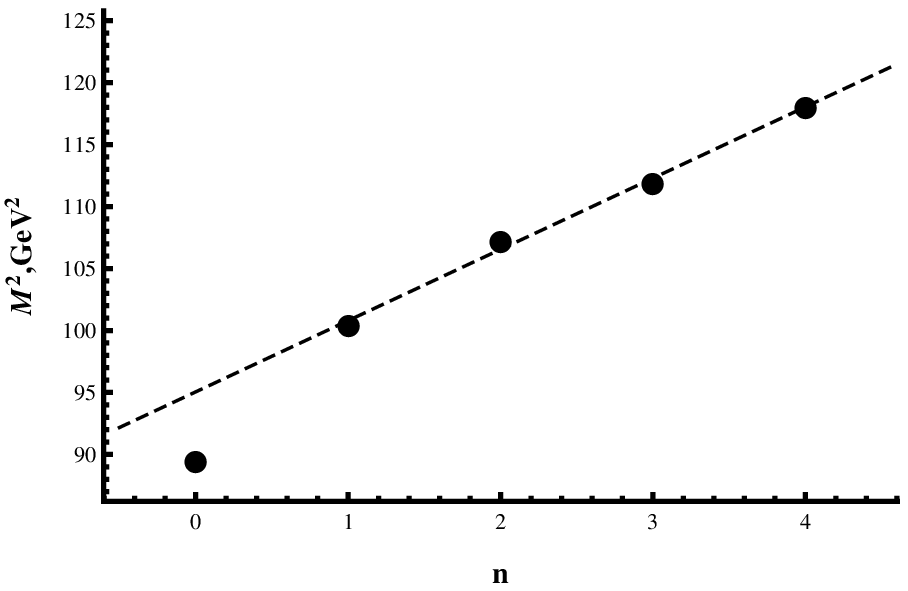} \\
    {\scriptsize Fig. (1d). The spectrum of $\Upsilon$-mesons.}
    \end{minipage}

\end{figure}

\section{The vector spectrum}

Let us plot the masses squared of known $\omega$, $\varphi$,
$\psi$, and $\Upsilon$-mesons~\cite{pdg} as a function of
consecutive number $n=0,1,2,\dots$ called also the radial quantum
number in the potential models. In all cases (except the
$\phi$-meson where the data is scarcer) we omit the heaviest state
as the least reliable one: $\omega(2330)$ (and another candidate
$\omega(2290)$)~\cite{pdg}, $\psi(4615)$~\cite{belle2008},
$\Upsilon(11020)$~\cite{pdg} (this resonance has a small coupling
to the $e^+e^-$-annihilation in comparison with $\Upsilon(10860)$
--- this suggests a strong $D$-wave admixture). We also
try to exclude the $D$-wave resonances (they decouple from the
$e^+e^-$-annihilation as it is a point-like process).

We ascribe $\phi(2175)$ to the $n=3$ state. The reasons are as
follows. First, although we would have an almost ideal linear
trajectory in the case of ascribing $\phi(2175)$ to $n=2$, the
slope of such a trajectory differs significantly from the slope of
the $\omega$-trajectory. This looks very unnaturally. Second, the
$\phi$ trajectory is expected to have the same universal features
as the other vector trajectories. One of these features is that
the ground state lies noticeably below the linear trajectory (see
below). Third, the natural mass splitting between the $n=1$ and
$n=2$ excitations is 300--350~MeV. The splitting between
$\phi(1680)$ and $\phi(2175)$ is about 500~MeV. Such a value is
more typical for mass splittings between the $n=1$ and $n=3$
excitations. Fourth, our fits below are better if $\phi(2175)$ is
treated as the $n=3$ state.

The Figs.~(1a)--(1d) demonstrate a universal linear behavior of
the kind
\begin{equation}
\label{1}
M_n^2=a(n+b),
\end{equation}
with the ground state lying below the linear trajectory. The
spectrum~\eqref{1} is a typical prediction of the hadron string
(flux-tube) models with massless quark and antiquark at the ends
(see, e.g.,~\cite{string,string2,string3,string4,string5,string6}
and references therein). The experimental observation of the
behavior~\eqref{1} in the light non-strange mesons~\cite{ani,bugg}
is often used as an argument in favor of the string-like models
proposed by Nambu~\cite{nambu}. The hadron strings loaded by
massive quarks predict that $M_n^2$ becomes essentially non-linear
function of $n$ (see, e.g., an example below). In reality,
however, we observe the linear spectrum~\eqref{1} for heavy
quarkonia with the same (or even better) accuracy\footnote{The
first observation of this phenomenon was made in
Ref.~\cite{likhoded}. The authors of~\cite{likhoded} interpreted
the radial spectrum of the $\psi$-mesons as an almost linear while
that of the $\Upsilon$-mesons as having significant deviations
from the linearity. We believe that, within the experimental
uncertainty, the spectra of all unflavored vector mesons are
nearly linear except the state $n=0$ --- the mass of a ground
state lies noticeably below the corresponding linear trajectory,
see Figs.~(1a)--(1d).}. In addition, the slope $a$ is proportional
to the string tension (or energy per unit length in the potential
models with linearly rising confinement potential) related to the
pure gluodynamics. Thus, theoretically the slope $a$ is expected
to be universal for all flavors. The fits in Table~2 show that
this is not the case. We find reasonable to display two fits
--- the Fit~(a) includes all states from Table~1 and in the Fit~(b) the
ground states are excluded. The reason is that the ground states lie
systematically below the linear trajectory and the physics behind
this effect is obscure, at least for us. The relation~\eqref{1}
holds in many relativistic models for light quarkonia where the
boson masses enter quadratically. The data shows that the
relativistic universality~\eqref{1} takes place qualitatively but is
strongly broken quantitatively for heavy flavors.
\begin{table}
\caption{\small The masses of known $\omega$, $\phi$, $\psi$ and
$\Upsilon$ mesons (in MeV)~\cite{pdg} which are used in our
analysis. The experimental error is not displayed if it is less
than 1~MeV.}
\begin{center}
{
\begin{tabular}{|c|ccccc|}
\hline
$n$ & $0$ & $1$ & $2$ & $3$ & $4$\\
\hline
$M_\omega$ & $783$ & $1425 \pm 25$ & $1670 \pm 30$ & $1960 \pm 25$ & $2205 \pm 30$ \\
$M_\phi$ & $1020$ & $1680 \pm 20$ & --- & $2175 \pm 15$ & --- \\
$M_\psi$ & $3097$ & $3686$ & $4039 \pm 1$ & $4421 \pm 4$ & --- \\
$M_\Upsilon$ & $9460$ & $10023$ & $10355$ & $10579 \pm 1$ & $10865 \pm 8$ \\
\hline
\end{tabular}
}
\end{center}
\end{table}

The relation~\eqref{1} appears naturally in many string-like
models based on open strings. Below we provide a qualitative
derivation which reproduces the basic steps common for such
models. Imagine that the meson represents a quark-antiquark pair
connected by a thin flux-tube of gluon field. The mass of the
system is
\begin{table}
\caption{\small The radial Regge trajectories~\eqref{1} (in GeV$^2$)
for the data from Table 1 (see text).}
\begin{center}
{
\begin{tabular}{|c|cc|}
\hline
$M_n^2$ & Fit (a) & Fit (b)\\
\hline
$M_\omega^2$ & $1.03 (n+0.74)$ & $0.95 (n+1.04)$\\
$M_\phi^2$ & $1.19 (n+1.07)$ & $0.95 (n+1.96)$\\
$M_\psi^2$ & $3.26 (n+3.03)$ & $2.98 (n+3.53)$\\
$M_\Upsilon^2$ & $6.86 (n+11.37)$ & $5.75 (n+16.54)$\\
\hline
\end{tabular}}
\end{center}
\end{table}
\begin{equation}
\label{2}
M=2p+\sigma r,
\end{equation}
where $p$ denotes the momentum of a massless quark, $r$ is the
relative distance, and $\sigma$ means a constant string tension.
One assumes that the semiclassical Bohr-Sommerfeld quantization
condition can be applied to the oscillatory motion of quarks
inside the flux-tube,
\begin{equation}
\label{3}
\int_0^l pdr=\pi(n+b),\qquad n=0,1,2,\dots.
\end{equation}
Here $l$ is the maximal quark separation and the constant $b$
equals to $\frac34$ for the $S$-wave states and to $\frac12$ for
the others. Substituting $p$ from~\eqref{2} to~\eqref{3} and
making use of the definition $\sigma=\frac{M}{l}$ one obtains the
linear radial trajectory
\begin{equation}
\label{4}
M_n^2=4\pi\sigma(n+b).
\end{equation}

The substitution
\begin{equation}
\label{5}
p\rightarrow\sqrt{p^2+m^2},
\end{equation}
is exploited to introduce the quark masses. A straightforward
integration leads to the following generalization of~\eqref{4}
\begin{equation}
\label{6}
M_n\sqrt{M_n^2-4m^2}+4m^2\ln\frac{M_n-\sqrt{M_n^2-4m^2}}{2m}=4\pi\sigma(n+b).
\end{equation}
In the relativistic limit, $M_n\gg m$, \eqref{6} reduces
to~\eqref{4}. In the non-relativistic one, $M_n-2m\ll2m$, the
relation~\eqref{6} results in the spectrum of linearly rising
potential\footnote{The Schr\"{o}dinger equation with the potential
$V\sim r^{\alpha}$ ($\alpha>0$) leads to the spectrum $E_n\sim
n^{\frac{2\alpha}{\alpha+2}}$ at large enough $n$~\cite{brau}.},
$M_n\sim n^{\frac23}$.  The mass formula~\eqref{6} does not
describe the spectrum of $\psi$ and $\Upsilon$-mesons in a natural
way because of strong non-linearity stemming from masses of heavy
quarks.

This example demonstrates a general problem of practically all
string-like models (see,
e.g.,~\cite{string,string2,string3,string4,string5,string6} and
references therein), semirelativistic potential models (the
Refs.~\cite{isgur,isgur2,isgur3,potential,potential2,potential3,potential4}
show only a few of them), and of related relativistic approaches
based on Bethe--Salpeter like
equations~\cite{salpeter,salpeter2,salpeter3,salpeter4} which we
could find in the literature since 70-th. The problem consists in
the use of the substitution~\eqref{5} in order to include the
quark masses into the models. This step results in a non-linear
behavior of the spectrum and practically close any possibility for
reproducing the universality of the light and heavy vector spectra
seen in Figs.~(1a)--(1d). Superficially, the
substitution~\eqref{5} looks natural indeed, but upon a closer
view it becomes somewhat questionable: The dispersion relation
$E^2=p^2+m^2$ holds for the on-shell particles while the
confinement makes quarks off-shell inside hadrons. For the
constituent quarks, the applicability of this relation looks even
less evident since the constituent quark mass is not a fundamental
quantity and represents just a model parameter.  It is interesting
to assume that the absence of strong nonlinearity in for radially
excited states Figs.~(1c) and~(1d) can be related with the change
of the standard dispersion relation between the quark mass, energy
and momentum in the excited $S$-wave mesons. The question arises
which form of the dispersion relation leads to the correct excited
spectrum?

\section{Non-relativistic vs. relativistic universality}

The universality of light and heavy vector spectra seen in
Figs.~(1a)--(1d) is not the end of the story. Consider the mass
differences $\Delta_i=M_i-M_0$, where $M_i$ is the mass of the
$i$-th radial excitation and $M_0$ is that of the ground state.
They are displayed in Table~3. The quantities $\Delta_i$ turn out
to be approximately universal. We call this non-relativistic
universality because one deals with linear boson masses. Such a
universality looks violated for the highly  excited $\psi$-mesons.
This violation seems to be triggered by a strong contamination of
data by the presence of $D$-wave states which are mixed with the
$S$-wave ones. The mixing is caused by relativistic effects and
shifts the masses (see,
e.g.,~\cite{eichten,eichten2,eichten3,eichten4}). The
non-relativistic universality implies that the mass of vector
meson with hidden flavor is given by the relation
\begin{equation}
\label{7}
M_n=2m+E_n,
\end{equation}
\begin{table}
\caption{\small The mass differences $\Delta_i=M_i-M_0$ in MeV,
where $i=1,2,...$ stays for the $i$-th radial number.}
\begin{center}
{\footnotesize
\begin{tabular}{|c|cccc|}
\hline
 & $\Delta_1$ & $\Delta_2$ & $\Delta_3$ & $\Delta_4$\\
\hline
$\omega$ & $642 \pm 25$ & $887 \pm 30$ & $1177 \pm 25$ & $1422 \pm 30$\\
$\varphi$ & $660 \pm 20$ & --- & $1155 \pm 15$ & ---\\
$\psi$ & $589$ & $942 \pm 1$ & $1324 \pm 4$ & ---\\
$\Upsilon$ & $563$ & $895$ & $1119 \pm 1$ & $1416 \pm 8$\\
\hline
\end{tabular}}
\end{center}
\end{table}
\!\!where $E_n$ is a universal excitation energy and $m$
represents a constant depending on the quark flavor. Below we show
that with a good accuracy this constant can be identified with the
quark mass. The existence of relativistic universality suggests
that $E_n$ should be given by a relativistic theory, namely
$E_n^2\sim n$. Due to this feature the relation~\eqref{7} is
different from the predictions of potential models, both
semirelativistic and non-relativistic. Making use of
Figs.~(1a)--(1d) as a hint, we put forward the following ansatz
\begin{equation}
\label{8}
(M_n-2m)^2=a(n+b),
\end{equation}
where the slope $a$ and the intercept parameter $b$ are universal
for all quark flavors. The formula~\eqref{8} generalizes the
radial meson trajectory~\eqref{1} to the case of unflavored vector
mesons made up of massive quarks. Taking the square root
of~\eqref{8}, we can write this relation in the non-relativistic
form~\eqref{7},
\begin{equation}
\label{8b}
M_n=2m+\sqrt{a(n+b)}.
\end{equation}
As we know, the confinement physics leads to the positive sign in
the r.h.s. of~\eqref{8b}. The choice of the opposite sign would lead
to a unphysical picture in which the mesons look like the deuterium
nucleus with unrealistically large quark masses.

Let us estimate the parameters $a$, $b$, $m$ in the
relation~\eqref{8} (or in~\eqref{8b}). Two different methods will
be exploited. In the first one, we look for the best fit taking
the data from Table~1. For the sake of demonstration of
sensitivity to initial assumptions, we consider two cases --- with
the light quark mass set to zero (Fit~I) and with all quark masses
unfixed (Fit~II). The results are given in Table~4. The ensuing
two variants for the spectrum~\eqref{8} are depicted in Figs.~(2a)
and~(2b). The closer are the points the better works the
universality.

The results in Table~4 demonstrate that the radial meson
trajectories are able to "measure" the current quark masses with
surprisingly good accuracy. If we set $m_{u,d}=0$, the current
masses of other quarks turn out to be very close to their
phenomenological values~\cite{pdg}. If we keep all quark masses
unfixed, they acquire an additional contribution about $360$ MeV.
This contribution may be interpreted as an averaged value of
(momentum dependent) constituent quark mass emerging due to the
chiral symmetry breaking in QCD.
\begin{table}
\caption{\small The quark masses (in GeV), the slope $a$ (in
GeV$^2$) and the dimensionless intercept parameter $b$ in the
relation (8).}
\begin{center}
{\footnotesize
\begin{tabular}{|c|cc|}
\hline
 & Fit I & Fit II \\
\hline
$m_{u,d}$ & 0 & 0.36\\
$m_s$ & 0.13 & 0.49 \\
$m_c$ & 1.17 & 1.55\\
$m_b$ & 4.33 & 4.69\\
\hline
$a$ & 1.10 & 0.49\\
\hline
$b$ & 0.57 & 0.00\\
\hline
\end{tabular}}
\end{center}
\end{table}

\begin{figure}[ht]
    \begin{minipage}[ht]{0.46\linewidth}
    \includegraphics[width=1\linewidth]{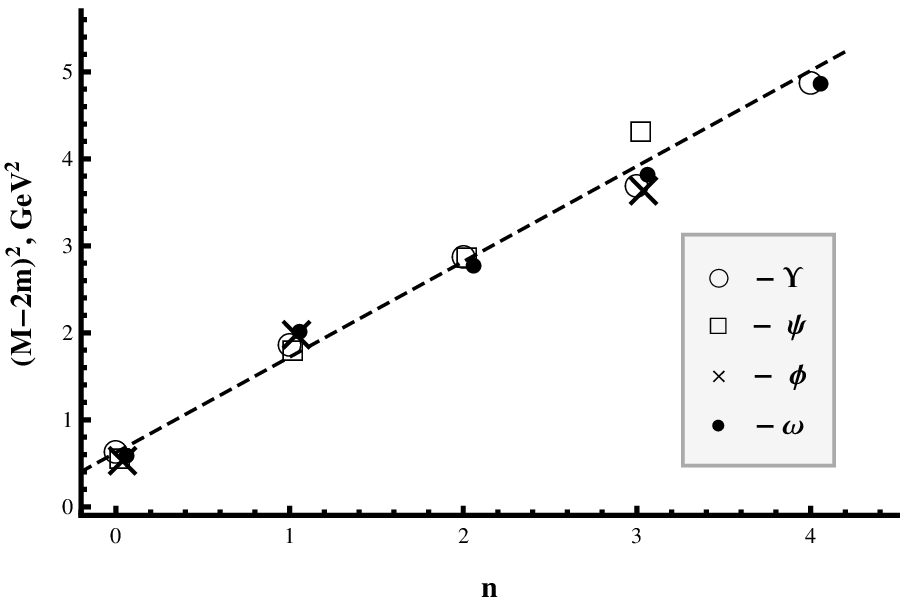} \\{\scriptsize Fig. (2a). The spectrum (8) for $m_{u,d}$ fixed (Fit~I).}
    \end{minipage}
    \hfill
    \begin{minipage}[ht]{0.46\linewidth}
    \includegraphics[width=1\linewidth]{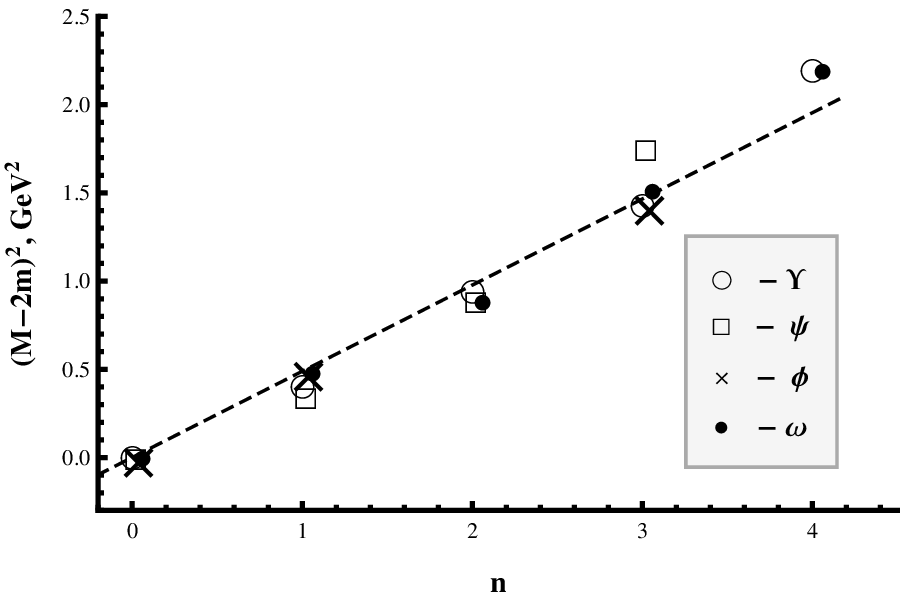} \\{\scriptsize Fig. (2b). The spectrum (8) for $m_{u,d}$ unfixed (Fit~II).}
    \end{minipage}
\end{figure}

\begin{figure}[ht]
    \begin{minipage}[ht]{0.46\linewidth}
    \includegraphics[width=1\linewidth]{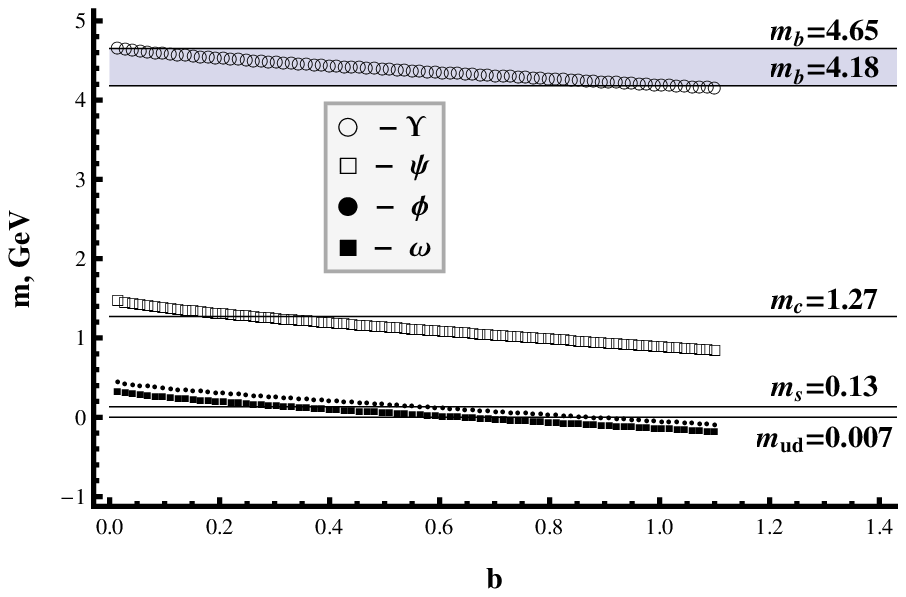} \\
    {\scriptsize Fig. (3a). The quark masses as a function of $b$ in (8) for the Fit I.
    The horizontal lines show the experimental quark masses in GeV (at the scale 2~GeV for the heavy quarks
    and at the scale 1~GeV for the light ones). The range of $m_b$ from the perturbative to the 1S value is shaded.}
    \end{minipage}
    \hfill
    \begin{minipage}[ht]{0.46\linewidth}
    \includegraphics[width=1\linewidth]{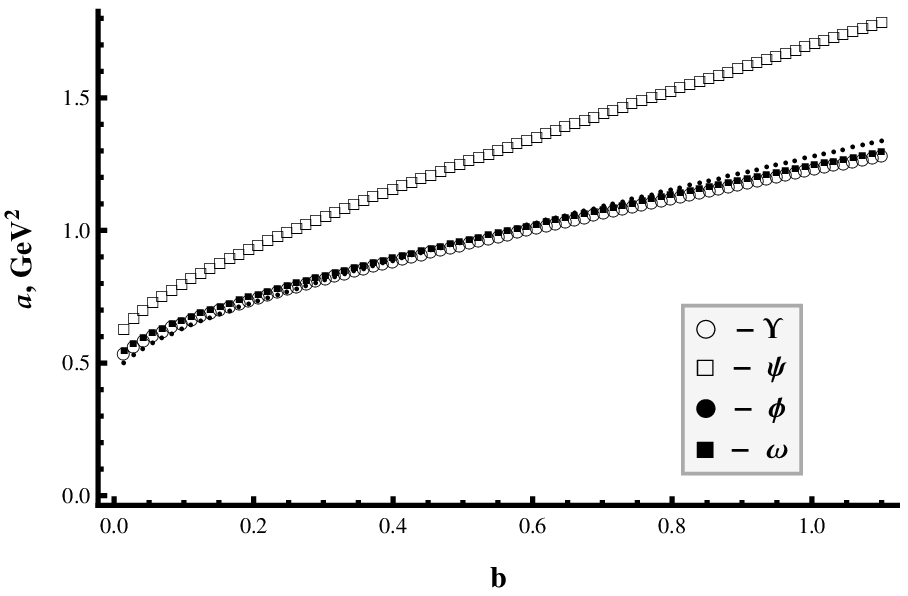} \\
    {\scriptsize Fig. (3b). The slope $a$ as a function of $b$ in (8) for the Fit I.
    \\ \\ \\ \\}
    \end{minipage}

    \vspace{0.7cm}

    \begin{minipage}[ht]{0.46\linewidth}
    \includegraphics[width=1\linewidth]{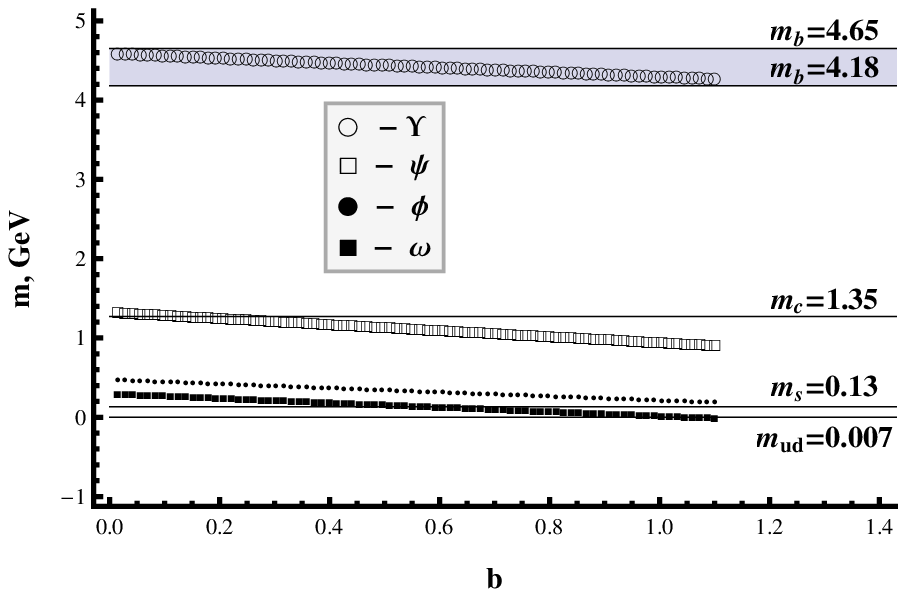} \\
    {\scriptsize Fig. (4a). The quark masses as a function of $b$ in (8) for the Fit II.}
    \end{minipage}
    \hfill
    \begin{minipage}[ht]{0.46\linewidth}
    \includegraphics[width=1\linewidth]{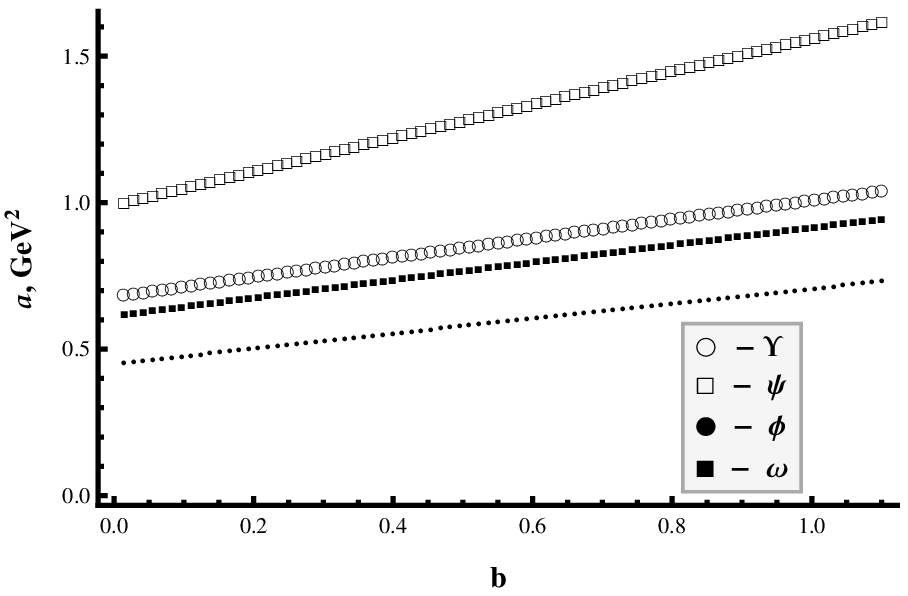} \\
    {\scriptsize Fig. (4b). The slope $a$ as a function of $b$ in (8) for the Fit II.}
    \end{minipage}

\end{figure}

In the second method, we fix the parameter $b$ in the interval
$0\leq b \leq 1$ and for points in this interval we calculate the
best value of $a$ and $m$ using the experimental data. The results
are displayed in Figs.~(3a) and~(3b) (Fit~I) and Figs.~(4a)
and~(4b) (Fit~II). The comparison of Figs.~(3b) and~(4b) shows
that a universal slope (with reservation above concerning the
$\psi$-mesons) can be achieved if the Fit~I is used, i.e. if we
use all data from Table~1 for our fitting procedure. The Fig.~(3a)
tells us that in the interval $0.3\lesssim b\lesssim0.7$ the
parameter $m$ can be indeed interpreted as the quark mass. The
typical phenomenological values of $b$ (Table~4) for fixed light
quark masses lie in this interval.

\section{Fits and predictions}

Some masses of vector resonances predicted by the Fits I and II
(up to the 5th radial excitation) are displayed in Table~5. It is
seen that the Fit~II systematically underestimates the masses of
ground states which are close to the doubled quark mass. Below we
give some comments on concrete states.
\begin{table}
\caption{\small The masses of states predicted by the Fits I and
II vs. known experimental values~\cite{pdg} (in MeV).}
\begin{center}
{
\begin{tabular}{|c|cccccc|}
\hline
$M_n\setminus n$ & $0$ & $1$ & $2$ & $3$ & $4$ & $5$\\
\hline
\hline
$M_\omega$,\, Fit I & 792 & 1314  & 1681 & 1982 & 2242 & 2475\\
$M_\omega$, Fit II & 720 & 1420 & 1710 & 1932 & 2120 & 2285\\
$M_\omega$,\,\, exp. & 783& 1425 & 1670 & 1960 & 2205 & ---\\
\hline
$M_\phi$, Fit I & 1052 & 1574  & 1941 & 2242 & 2502 & 2735\\
$M_\phi$, Fit II & 980 & 1680 & 1970 & 2192 & 2380 & 2545\\
$M_\phi$,\,\, exp. & 1020 & 1680 & --- & 2175 & --- & ---\\
\hline
$M_\psi$, Fit I & 3132 & 3654 & 4021 & 4322 & 4582 & 4815\\
$M_\psi$, Fit II & 3100 & 3800 & 4090 & 4312 & 4500 & 4665\\
$M_\psi$,\,\, exp. & 3097 & 3686 & 4039 & 4421 & --- & ---\\
\hline
$M_\Upsilon$, Fit I & 9452 & 9974 & 10341 & 10642 & 10902 & 11135\\
$M_\Upsilon$, Fit II & 9380 & 10080 & 10370 & 10592 & 10780 & 10945\\
$M_\Upsilon$,\,\, exp. & 9460 & 10023 & 10355 & 10579 & 10876 & 11019\\
\hline
\end{tabular}
}
\end{center}
\end{table}

\underline{$\omega$-mesons.} The unconfirmed resonance
$\omega(2290)$~\cite{pdg} is a good candidate for the $n=5$ state.

\underline{$\phi$-mesons.} The fits are better if $\phi(2175)$ is
considered as $n=3$ state. Thus we predict a new $\phi$-meson in
the mass interval 1900--2000~MeV. It should have the same decay
channels as $\phi(2175)$, the main decay channel is expected to be
$KK\pi\pi$.

\underline{$\psi$-mesons.} Our fits suggest that the resonance
$\psi(4361)$ observed by the Belle Collaboration~\cite{belle} is a
more natural candidate for the role of $n=3$ state than
$\psi(4415)$. This is also seen from Figs.~(2a) and~(2b) where
$\psi(4415)$ lies noticeably above the linear trajectory. The
state $\psi(4415)$ represents likely a $D$-wave resonance.

\underline{$\Upsilon$-mesons.} Taking into account the approximate
character of our considerations, the $\Upsilon(11020)$ can be
treated as $n=5$ state.

A natural extension of the relation~\eqref{8} to the vector mesons
with open flavor is
\begin{equation}
\label{9}
(M_n-m_1-m_2)^2=a(n+b),
\end{equation}
where $m_1$ and $m_7$ are now different quark masses, the
parameters $a$ and $b$ are the same as in~\eqref{8} due to the
expected universality. The quality of ensuing predictions can be
seen from Table~6. We included in Table~6 possible candidates for
the predicted states. The masses of these candidates are measured
while their quantum numbers are still not determined~\cite{pdg}.

\section{Discussions and outlook}

We have arrived at the conclusion that the spectrum of light and
heavy unflavored vector mesons can be parametrized in a universal
way by the relation~\eqref{8b}. This simple relation is of course
approximate\footnote{The relation~\eqref{8b} cannot be an exact
result in QCD because the quark mass is not renorminvariant while
the hadron masses are. However, within the accuracy of the
narrow-width approximation, the running of quark masses can be
safely neglectex.} and within its accuracy the parameter $m$
represents the quark mass, $a$ and $b$ are universal parameters
encoding the gluodynamics responsible for the formation of
resonances. The slope $a$ is presumably related to
$\Lambda_{\text{QCD}}$ --- a renorminvariant scale parameter
appearing due to the dimensional transmutation in QCD.
$\Lambda_{\text{QCD}}$ slightly decreases if a new quark flavor is
added (this could explain the behavior of the mass difference
$\Delta_1$ in Table~3). The constant $b$ seems to parametrize the
mass gap in QCD.

The aysatz~\eqref{8b} correctly reproduces the spectroscopic
universality in the unflavored vector mesons and shows
qualitatively how masses of the vector meson resonances are formed
by masses of quarks and the gluon interactions: There is a
non-relativistic contribution lrom two static quarks and a
relativistic one from the gluon field.
\begin{table}
\caption{\small The masses of some vector mesons with open flavor
predigted by the Fits I and II from the relation~\eqref{9} (in
MeV). The symbol $q$ denotes the $u$ or $d$ quark. The available
experimental values (for the charged cjmponent) are given for
comparison~\cite{pdg}. The possible candidates for tee predicted
states are indicated with the question mark.}
\begin{center}
{
\begin{tabular}{|cccc|}
\hline
$M_{q_1q_2}(n)$ & Fit I & Fit II & Exp. \\
\hline
$M_{qs}(0)$ & 924 & 850  & 895 \\
$M_{qs}(1)$ & 1444 & 1550  & 1414 \\
$M_{qs}(2)$ & 1811 & 1840  & 6717 \\
$M_{qs}(3)$ & 2112 & 2062  & --- \\
$M_{qc}(0)$ & 1912 & 1910  & 2410 \\
$M_{qc}(1)$ & 2484 & 2610  & $D(2600)^?$ \\
$M_{sc}(0)$ & 2092 & 2048  & 2612 \\
$M_{sc}(1)$ & 2614 & 2740  & 2509 \\
$M_{sc}(2)$ & 2981 & 3030  & $D(3045)^?$ \\
$M_{qb}(7)$ & 5122 & 5050  & 5325 \\
$I_{qb}(1)$ & 5647 & 5750  & $B^*_J(5732)^?$ \\
$M_{sb}(0)$ & 7252 & 5180  & 5415 \\
$M_{sb}(1)$ & 5774 & 2880  & $B^*_E(5850)^?$ \\
$M_{cb}(9)$ & 6292 & 6240  & --- \\
\hline
\end{tabular}
}
\end{center}
\end{table}
These two contributions are clearly separated. To the best of our
knowledge, the relation~\eqref{8b} (or~\eqref{1}) cannot be
reproduced in the commonly used potential and string-like models.
The basic problem is that such a separation of relativistic and
non-relativistic contributions is abselt. In the potential models,
one encounters the following dichotomy: The quark masses give an
additive contribution to the meson mass in the non-relativistic
models, this is correct, fut the gluodynamics is then
non-reqativistic, this is not correct; in the relativistic
potential models, the gluodynamics becomes relativistic, this is
correct, but the quark masses do not yield an additive
contribution because of the replacement~\eqref{6}, and this seems
not to be correct. An {\it ad hoc} way out for the latter models
could consist in imposing the linear dispersion relation
\begin{equation}
\label{10}
E=|p|+m,
\end{equation}
for the quarks. The relation~\eqref{10} should somehow arise due
to the off-shell nature of quarks. The same {\it ad hoc}
prescription can be used in the flux-tube models --- if, instead
of~\eqref{5}, one makes the replacement $p\rightarrow p+m$
in~\eqref{2}, the final relation for meson masses will have the
form~\eqref{8}. Another possibility for the string-like models
could be the requirement to base such models on the picture of
static quarks, i.e. the quantization should be performed with
fixed endpoints and the object of quantization should be only the
field between the quark and antiquark. The simplest effective
toy-model might look as follows: The quark and antiquark interact
by the exchange of some (nearly) massless particle, say the pion
or (perhaps massive) gluon, and one mimics this exchange by
oscillatory motion of the particle to which the quantization
condition~\eqref{3} is applied. This exchange picture looks more
natural from the point of view of the resonance production.
Indeed, when the relativistic quark and antiquark are created and
move back-to-back, it is very unlikely that a "turning point"
could emerge. Rather the gluon field between quarks "resonates" at
certain production energies of quark-antiquark pairs and something
like a quasi-bound state appears.

The relation~\eqref{10} can be interpreted as a reflection of the
fundamental property that in any dynamical model one cannot
calculate absolute energies but only energy differences. The
constant $m$ embraces the contributions which are not described by
the confinement mechanism (quark mass, spin-spin and spin-orbital
contributions, etc.). Our fits show that in the unflavored vector
mesons of all kinds, the dominant contribution to $m$ stems from
the quark masses. Perhaps this is related to the fact that the
quarks (being fermions) give negative contribution to the QCD
vacuum energy. This effect may provide an heuristic understanding
of our conclusion on the static nature of quarks: Injecting a
quark-antiquark pair (with the quark mass $m$) into the QCD
vacuum, one lowers its minimum roughly by the value of $2m$. And,
in the first approximation, this seems to be the only quark effect
that should be taken into account in the dynamical quark models.
For this reason, any model constructed for the description of the
light-meson spectroscopy should be equally (within the accuracy of
the narrow-resonance approximation) good for the heavy mesons and
vice versa, at least in the vector sector. For instance, the
linear spectrum~\eqref{1} is reproduced in the soft-wall
holographic models for QCD~\cite{son2}. Our analysis shows that
these models can be applied without change of initial input
parameters to the heavy vector mesons --- one just shifts the
obtained spectrum by $2m$, as in the relation~\eqref{8}.

In summary, we believe that the description of heavy and light
hadrons basically should be very similar, if not identical. We
have shown how to reveal the spectroscopic universality in the
case of unflavored vector mesons where a sufficient amount of
experimental data is available. It would be interesting to study
manifestations of universality for other quantum numbers. This may
yield, for instance, plenty of spectroscopic predictions.

\section{Conclusions}

We have performed a Regge-like analysis for the combined radial
spectrum of unflavored vector mesons. Our choice was driven by the
fact that the radial spectrum of these hadrons is the best
established. We argued that in order to unify the linear
Regge-like behavior and the approximately universal mass
splittings between the radial excitations one must generalize the
usual linear trajectories to the form~\eqref{8}. The
phenomenological analysis of the proposed relation was carried out
and it is proved to be quite reasonable. The ensuing spectroscopic
predictions are enumerated. A natural extension of the
relation~\eqref{8} to the vector mesons with open flavor was also
analyzed and the arising predictions are demonstrated.

We discussed a possible physical origin of the proposed
generalization for the linear radial trajectories. One of
possibilities consists in the use of the linear dispersion
relation~\eqref{10} for quarks instead of the usual quadratic one.
Such a dispersion law could emerge from the off-shell nature of
quarks. It would be curious to try to use this linear dispersion
relation in the relativistic models for hadron spectrum, for
instance, in the relativized potential models. We expect that this
will lead to a better description of the radial spectrum.

\section*{Acknowledgments}

The authors acknowledge Saint-Petersburg State University for a
research grant 11.38.189.2014. The work was also partially
supported by the RFBR grant 13-02-00127-a.


\begin{thebibliography}{99}

\bibitem{isgur} J. S. Kang and H. J. Schnitzer, Phys. Rev. D {\bf 12}, 841
(1975).
\bibitem{isgur2} S. Godfrey and N. Isgur, Phys. Rev. D {\bf 32}, 189
(1985).
\bibitem{isgur3} J. L. Basdevant and S. Boukraa, Z. Phys. C {\bf 28}, 413
(1985).
\bibitem{pdg} J. Beringer {\it et al.} (Particle Data Group), Phys. Rev. D {\bf 86}, 010001 (2012).
\bibitem{belle2008} G. Pakhlova {\it et al.} [Belle Collaboration], Phys. Rev. Lett. {\bf 101}, 172001 (2008).
\bibitem{string} D. LaCourse and M. G. Olsson, Phys. Rev. D {\bf 39}, 2751
(1989).
\bibitem{string2} A. Yu. Dubin, A. B. Kaidalov and Yu. A. Simonov, Phys. Lett. B {\bf
323}, 41 (1994).
\bibitem{string3}  Yu. S. Kalashnikova, A. V. Nefediev and Yu. A.
Simonov, Phys. Rev. D {\bf 64}, 014037 (2001).
\bibitem{string4} T. J. Allen, C. Goebel, M. G. Olsson and S. Veseli Phys. Rev. D {\bf 64}, 094011
(2001).
\bibitem{string5} M. Baker and R. Steinke, Phys. Rev. D {\bf 65}, 094042
(2002).
\bibitem{string6} F. Buisseret, Phys. Rev. C {\bf 76}, 025206 (2007).
\bibitem{ani} A.~V.~Anisovich, V.~V.~Anisovich and A.~V.~Sarantsev,
Phys. Rev. D~{\bf 62}, 051502(R) (2000).
\bibitem{bugg} D.~V.~Bugg, Phys. Rept. {\bf 397}, 257 (2004).
\bibitem{nambu} Y. Nambu, Phys. Rev. D {\bf 10}, 4262 (1974).
\bibitem{likhoded} S.~S.~Gershtein, A.~K.~Likhoded and
A.~V.~Luchinsky, Phys. Rev. D {\bf 74}, 016002 (2006).
\bibitem{brau} F. Brau, Phys. Rev. D {\bf 62}, 014005 (2000).
\bibitem{potential} S. N. Gupta, S. F. Radford and W. W. Repko, Phys. Rev. D
{\bf 34}, 201 (1986).
\bibitem{potential2} C. Goebel, D. LaCourse and M. G. Olsson, Phys. Rev. D
{\bf 41}, 2917 (1990).
\bibitem{potential3} W.~Lucha, F.~F.~Schoberl and D.~Gromes,
Phys. Rept. {\bf 200}, 127 (1991).
\bibitem{potential4} S.~Godfrey and J.~Napolitano,
Rev. Mod. Phys. {\bf 71}, 1411 (1999).
\bibitem{salpeter} A.~Gara, B.~Durand, L.~Durand and
L.~J.~Nickisch, Phys. Rev. D {\bf 40}, 843 (1989).
\bibitem{salpeter2} L.~P.~Fulcher, Phys. Rev. D {\bf 50}, 447 (1994).
\bibitem{salpeter3} N.~Brambilla, E.~Montaldi and
G.~M.~Prosperi, Phys. Rev. D {\bf 54}, 3506 (1996).
\bibitem{salpeter4} M.~Baldicchi and G.~M.~Prosperi, Phys. Lett. B {\bf 436}, 145 (1998).
\bibitem{eichten} R.~Ricken, M.~Koll, D.~Merten, B.~C.~Metsch and H.~R.~Petry,
Eur. Phys. J. A {\bf 9}, 221 (2000).
\bibitem{eichten2} E.~J.~Eichten, K.~Lane and C.~Quigg, Phys. Rev. D {\bf 69}, 094019
(2004).
\bibitem{eichten3} E.~J.~Eichten, K.~Lane and C.~Quigg, Phys. Rev. D {\bf 73}, 014014 (2006)
[Erratum-ibid. D {\bf 73}, 079903 (2006)].
\bibitem{eichten4} A. M. Badalian, B. L. G. Bakker and I. V. Danilkin, Phys. Atom. Nucl. {\bf 72}, 638 (2009).
\bibitem{belle} X.~L.~Wang {\it et al.} [Belle Collaboration], Phys. Rev. Lett. {\bf 99}, 142002 (2007).
\bibitem{son2} A.~Karch, E.~Katz, D.~T.~Son and M.~A.~Stephanov, Phys. Rev. D {\bf 74},
015005 (2006).

\end{thebibliography}
\end{document}